\documentclass[%
 reprint,
superscriptaddress,
amsmath,amssymb,
aps,
floatfix,
]{revtex4-2}

\usepackage{graphicx}
\usepackage{dcolumn}
\usepackage{bm}
\usepackage[colorlinks]{hyperref}

\usepackage{float}
\usepackage{makecell}
\usepackage{comment}
\usepackage{enumitem}
\usepackage[thinlines]{easytable}
\usepackage{algorithm}
\usepackage[noend]{algpseudocode}

\makeatletter
\def\ftype@algorithm{8}
\makeatother

\usepackage{hhline}
\usepackage{braket, siunitx, placeins}
\usepackage{epsfig}
\usepackage{siunitx}
\begin{document}

\preprint{APS/123-QED}

\title{Hardware-efficient erasure-error detection with an integer fluxonium}

\def\RLEaffil{Research Laboratory of Electronics, Massachusetts Institute of Technology, Cambridge, Massachusetts 02139, USA}
\def\LLaffil{Lincoln Laboratory, Massachusetts Institute of Technology, Lexington, Massachusetts 02421, USA}
\def\Physaffil{Department of Physics, Massachusetts Institute of Technology, Cambridge, Massachusetts 02139, USA}
\def\EECSaffil{Department of Electrical Engineering and Computer Science, Massachusetts Institute of Technology, Cambridge, Massachusetts 02139, USA}
\def\affilIBM{IBM Quantum, IBM Research Cambridge, Cambridge, Massachusetts 02142}
\def\affilNYU{Department of Physics, New York University, New York, New York 10003}

\newcommand{\fref}[1]{Fig.~\ref{#1}}
\newcommand{\eref}[1]{Eq.~(\ref{#1})}
\newcommand{\aref}[1]{Appendix~\ref{#1}}
\newcommand{\tref}[1]{Table~\ref{#1}}

\author{Junyoung~An}
\affiliation{\RLEaffil}
\affiliation{\EECSaffil}

\author{Helin~Zhang}
\affiliation{\RLEaffil}

\author{Jeffrey~M.~Gertler}
\affiliation{\LLaffil}

\author{Kate~Azar}
\affiliation{\RLEaffil}

\author{Ren\'ee~DePencier~Pi\~nero}
\affiliation{\LLaffil}

\author{Michael~Gingras}
\affiliation{\LLaffil}

\author{Junghyun~Kim}
\affiliation{\RLEaffil}
\affiliation{\EECSaffil}

\author{Bethany~M.~Niedzielski}
\affiliation{\LLaffil}

\author{Ilan~T.~Rosen}
\altaffiliation[Present address: ]{\affilIBM}
\affiliation{\RLEaffil}

\author{Mollie~E.~Schwartz}
\affiliation{\LLaffil}

\author{Joel~\^I-j.~Wang}
\altaffiliation[Present address: ]{\affilNYU}
\affiliation{\RLEaffil}

\author{Terry~P.~Orlando}
\affiliation{\RLEaffil}
\affiliation{\EECSaffil}

\author{Jeffrey~A.~Grover}
\affiliation{\RLEaffil}

\author{Max~Hays}
\affiliation{\RLEaffil}

\author{Kyle~Serniak}
\affiliation{\RLEaffil}
\affiliation{\LLaffil}

\author{William~D.~Oliver}
\email{william.oliver@mit.edu}
\affiliation{\RLEaffil}
\affiliation{\EECSaffil}
\affiliation{\Physaffil}

\date{\today}

\begin{abstract}

Erasure-error detection can improve the efficiency of quantum error correction by revealing the times and locations of their error events. In this work, we demonstrate erasure conversions and mid-circuit erasure detections in a single integer fluxonium, in which the states $\mathrm{|g\rangle, |f\rangle}$ encode the logical states and $\mathrm{|e\rangle}$ encodes the erasure state. The integer fluxonium suppresses direct $|\mathrm{f} \rangle \rightarrow |\mathrm{g} \rangle$ transitions and allows the dominant $|\mathrm{f} \rangle \rightarrow |\mathrm{e}\rangle$ transitions to be converted into detectable erasures. Furthermore, we identified a design space that nullifies the resonant-frequency shift between the two logical states, enabling ancilla-free mid-circuit erasure checks using the same resonator employed for final readout. By discarding the detected erasure events, we achieved an 8.4-fold increase in the $|\mathrm{f}\rangle$ state lifetime, a 1.38-fold increase in the Hahn-echo time, and a reduction of single-qubit gate error from 0.061(2)\% to 0.030(5)\%. Our results establish integer fluxonium as a hardware-efficient platform for erasure-error detection and conversion, while identifying the improvements required to realize an effective erasure qubit with high erasure bias.

\end{abstract}
\maketitle


\section{\label{sec:introduction} INTRODUCTION}
Quantum error correction (QEC) is widely regarded as a primary pathway toward fault-tolerant quantum computation. Provided the physical error rate remains below a specific error threshold, the logical error rate can be suppressed exponentially with increasing code distance~\cite{Preskill1998_qecthreshold, Willow2025, Bravyi2013_qec_scaling, Fowler2013_qec_scaling, Watson2014_qec_scaling, Fowler_surfacecode_scaling}. The magnitude of the error threshold depends on the underlying error model of the system. Specifically, erasure errors~\cite{Grassl1997_erasure_code, Bennett1997_erasure_channel} — error events whose timing and location are identified via direct measurement—allow significantly higher error thresholds compared to Pauli errors~\cite{Dumer2015_eraserror_benefit, Knill2005_eraserror_benefit, Stace2009_eraserror_benefit, Barrett2010_eraserror_benefit, Colmenarez2025_eraserror_benefit, ShouzhenGu2025_eraserror_benefit, Sahay2023_eraserror_benefit_neutralatom, Kubica2023_erasure_T1overcome, Gu2025_eraserr_simulation}. 

However, leveraging the advantages of erasure errors is often non-trivial, as they are not the dominant decoherence mechanism in many systems. In conventional superconducting qubit architectures that utilize the first two energy levels for computation, leakage out of the computational subspace is typically tiny compared to the Pauli errors and further suppressed by various techniques ~\cite{Acharya2023_Sycamore, Chen2016_leakage_suppress, Chiaro2025_leakage_cancel, Xin2025_leakage_reduction, Yang2024_leakage_reduction, Lacroix2025_leakage_reduction, Marques2023_mwleakage_reduction, Chiaro2025_leakage_cancel, Chiaro2025_leakage_cancel} such as pulse engineering or the use of leakage reduction units. To take advantage of the erasure error model, the hardware should be specifically engineered to ensure that erasure errors constitute the primary error channel while the Pauli errors are intrinsically suppressed; a system designed with this hierarchy is referred to as an \textit{erasure qubit}. This asymmetry between erasure errors and Pauli errors is known as the \textit{erasure-bias} condition.

A typical approach to implementing erasure qubits in superconducting circuit hardware is by using a dual-rail architecture~\cite{Shim2016_dualrail}. Dual-rail architecture uses two physical qubits~\cite{Campbell2020_transmondr, Levine2024_awstransmondr, Huang2026_iswaptransmondr, Hung2026_transmondr_chimatch} or cavities~\cite{Chou2024_cavitydr, deGraaf2025_cavitydr, Koottandavia2024_cavitydr, Mehta2025_dwavecavitydr} to create a computational subspace composed of a single excitation in each component, and the global ground state serves as an erasure state. Due to their energy-level configurations, the erasure-bias condition can be naturally met by dual-rail qubits, making them excellent candidates for erasure qubits. However, dual-rail architectures increase the hardware demand as they require two physical qubits to create one erasure qubit, and often an additional ancillary qubit which is dedicated to the erasure check. To reduce the hardware requirements, ideas that utilize three levels within a single superconducting qubit (qudit) have been proposed, such as a g-f erasure qubit~\cite{ChenluLiu2026_gfflxnm, JiashengMai2026_gfcavity, Wang2026_gfflxnmerasure, BaojieLiu2026_transmongf_eras_qubit} implemented using transmons or fluxoniums.

In this paper, we investigated integer fluxonium~\cite{Ardati2024_intflxnm, Mencia2024_intflxnm, Wang2025_Integer_fluxonium_CZgate} as a platform for erasure conversion and direct erasure detection. In an integer fluxonium, the $|\mathrm{f} \rangle \rightarrow |\mathrm{e}\rangle$ relaxation rate can be substantially stronger than the direct $|\mathrm{f}\rangle \rightarrow |\mathrm{g} \rangle$ relaxation, allowing a high fraction of relaxation events to be detected by encoding $|\mathrm{e}\rangle$ as an erasure state. In addition, the qubit-resonator system can be designed such that the readout resonator remains nearly insensitive to the logical states, which enables direct, ancilla-free erasure checks. We experimentally verified these properties by achieving the conditional improvements of coherence time and gate errors after postselecting against the detected erasure events. Furthermore, we discuss limitations of our current implementations such as missed erasures or additional errors induced by the erasure checks.

\section{\label{sec:device_working_principle} DEVICE WORKING PRINCIPLE}

\subsection{g-f erasure qubit}

A g-f erasure qubit is encoded such that the ground state $|\mathrm{g} \rangle $ and the second excited state $|\mathrm{f} \rangle$ form the computational subspace, while the first excited state $|\mathrm{e} \rangle $ serves as the erasure state [\fref{fig:device_and_schematics}(a)]. Thus, mid-circuit erasure checks are performed by probing the erasure state $|\mathrm{e}\rangle $ and flagging an error if population is detected there. 

To qualify as an effective erasure qubit, erasure events, the energy relaxation from $|\mathrm{f}\rangle$ to $\mathrm{|e\rangle}$, should dominate over residual Pauli errors such as bit-flips and phase-flips within the computational subspace. Furthermore, any additional dephasing induced from the mid-circuit erasure checks should be minimized.

In the following paragraphs, we introduce properties of an integer fluxonium that make it a suitable candidate for a g-f erasure qubit: decoherence protection offered across its first three energy levels and the capability to suppress the logical-state dispersive shift compared to the erasure state ($|\chi_{\mathrm{gf}}|\ll |\chi_{\mathrm{ge}}|$).

\subsection{Noise protection for an effective erasure conversion}

\begin{figure}[t]
    \centering
    \includegraphics[width=0.99\linewidth]{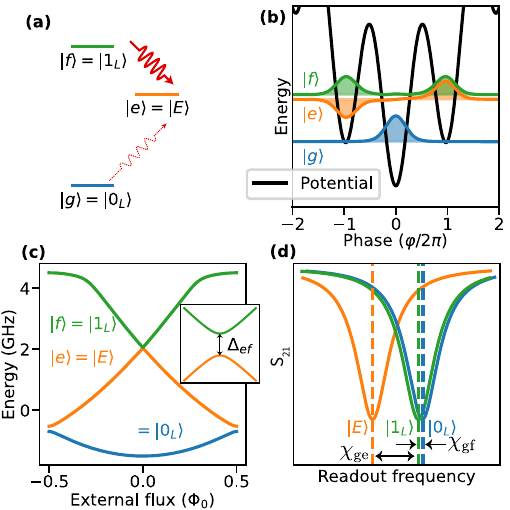}
    \caption{(a) Concept of a g-f erasure qubit. The computational subspace is spanned by $\mathrm{|g}\rangle = |0_{L}\rangle$ and $\mathrm{|f \rangle} = |1_{L} \rangle$, and the first excited state $|\mathrm{e} \rangle $ becomes the erasure state $|E\rangle$. The primary error channel is the $|\mathrm{f}\rangle \rightarrow|\mathrm{e} \rangle$ transition, while the errors within the computational subspace are suppressed. (b) Phase-space wavefunctions of the first three energy eigenstates from an integer fluxonium. States $|\mathrm{e}\rangle$ and $|\mathrm{f} \rangle$ form a symmetric-antisymmetric pair across the two fluxon wells of phase values of approximately $\pm 2\pi$. (c) First three energy levels of an integer fluxonium, centered at zero flux. At zero flux, the e-f transition frequency ($\Delta_{\mathrm{ef}}$) is reduced, typically ranging from 10--\SI{100}{MHz} (inset). (d) Desired dispersive shift configuration of the readout resonator of the g-f erasure qubit. The resonant frequency shift between the two logical states ($\chi_{\mathrm{gf}}$) should be minimized while the shift for the erasure state $|E\rangle$ is significantly larger to minimize backaction from the mid-circuit erasure checks.}
    \label{fig:device_and_schematics}
\end{figure}

Integer fluxonium is defined as a fluxonium biased at zero flux and with an inductive energy $E_{L}$ satisfying the condition $2\pi^{2} E_{L} \ll \sqrt{8 E_{J} E_{C}}$~\cite{Ardati2024_intflxnm}. In this parameter regime, the first and second excited states form a symmetric-antisymmetric pair across the two fluxon wells centered at $\varphi/2\pi=\pm1$ [\fref{fig:device_and_schematics}(b)], resulting in a three-level system characterized by $\omega_{\mathrm{ge}} \gg \omega_{\mathrm{ef}}$ at zero flux [\fref{fig:device_and_schematics}(c)]. This specific energy landscape results in the following protections for the three-level system consisting of $\mathrm{|g \rangle}$, $\mathrm{|e \rangle}$, and $\mathrm{|f \rangle}$. 

\begin{itemize}[leftmargin=*]
    \item The one-photon $\mathrm{|g \rangle \leftrightarrow |f \rangle}$ transition is forbidden by the parity-selection rule at zero flux. This makes the states $\mathrm{|g\rangle}$ and $\mathrm{|f\rangle}$ natural candidates for computational states.
    \item The $\mathrm{|g \rangle \leftrightarrow |e \rangle}$ transition is suppressed by a reduced matrix element, arising from the disjoint support of the two wavefunctions in phase space. This suppression allows a long-lived erasure state.
    \item The $\mathrm{|e \rangle \leftrightarrow |f \rangle}$ transition is suppressed by the small bi-fluxon tunneling amplitude across the displaced fluxon wells centered at $\varphi/2\pi = \pm 1$. This property contributes to the erasure-state lifetime and also the low overall erasure error rate.
\end{itemize}

Furthermore, the low-frequency dephasing within the computational subspace is effectively suppressed by operating at the zero-flux sweet spot. The protection from flux noise is additionally aided by a relatively small $E_{L}$, which typically ranges from 0.1--\SI{0.3}{GHz}. Any residual dephasing can be further mitigated through the application of dynamical decoupling techniques. The shot-noise dephasing is inherently mitigated, as minimizing the logical-state dispersive shift ($\chi_{\mathrm{gf}}$) is already a prerequisite to minimize dephasing during the mid-circuit erasure check.
We also note that an erasure check has a non-unitary back-action on the $\mathrm{g/f}$ manifold even with $\mathrm{\chi_{gf}} = 0$, because lack of a $|\mathrm{f\rangle\rightarrow |e\rangle}$ decay event indicates that the system is more likely to be in $\mathrm{|g\rangle}$. 
However, this effect does not enter into the measurements we present below because these dynamics conserve state purity and can be echoed away. 

\subsection{Suppressing erasure-check-induced dephasing} 

Another crucial property of an integer fluxonium as a g-f erasure qubit is the ability to nullify the resonant frequency shift between the logical states ($\chi_{\mathrm{gf}}$) [\fref{fig:device_and_schematics}(d)]. We have identified a parameter space where $\chi_{\mathrm{gf}}$ is canceled while $\chi_{\mathrm{ge}}$ remains on the order of several megahertz. This configuration allows high-fidelity mid-circuit erasure checks via the coupled readout resonator without introducing substantial measurement-induced dephasing to the logical states. Consequently, this capability alleviates the need for an additional qubit for erasure detection. A more detailed discussion on the parameter selection is in \aref{app:parameter_regime}. 

In summary, the integer fluxonium exhibits an inherently suppressed error rate within the computational subspace and can also provide intrinsic protection against measurement-induced dephasing during direct, ancilla-free erasure checks. In the next section, we present the experimental characterization of our integer fluxonium and evaulate erasure-detection capability.

\section{Measurement Results}
\subsection{\label{sec:basic_characterization} Basic characterization of the device}

\begin{figure}[t]
    \centering
    \includegraphics[width=0.99\linewidth]{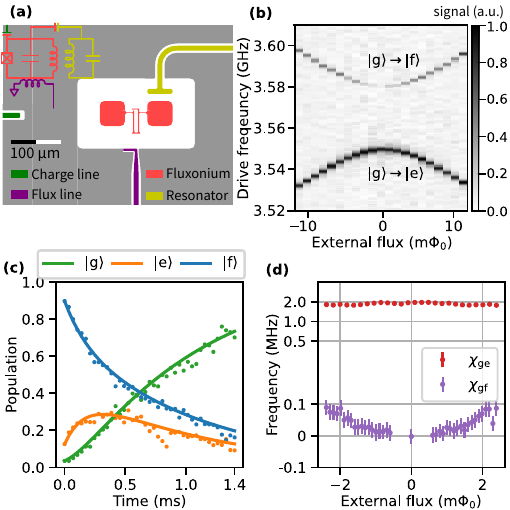}
    \caption{(a) Device geometry of the integer fluxonium qubit. The qubit is coupled to a charge line, flux line, and a readout resonator. (b) Qubit spectrum near zero flux. We calibrated a g-e transition and g-f transition frequencies. Notably, the g-f transition disappears at zero flux as expected by the parity-selection rule. The signal magnitude is rescaled such that the minimum and maximum signal magnitudes are normalized to zero and one. (c) We measured population evolution of the $\mathrm{|g\rangle,|e\rangle,|f\rangle}$ states after initialization into $\mathrm{| f\rangle}$. Solid lines represent numerical fits to a three-state energy relaxation model to extract $T_{1}^{\mathrm{f \rightarrow e}}$, $T_{1}^{\mathrm{e \rightarrow g}}$. (d) We measured dispersive shifts $\chi_{\mathrm{gf}}$, $\chi_{\mathrm{ge}}$ near zero flux. At zero flux, we obtained $\chi_{\mathrm{ge}}/2\pi=${1.86}$\pm$\SI{0.03}{MHz} and $\chi_{\mathrm{gf}}/2\pi=$-11.9$\pm$\SI{13.8}{kHz}.}
    \label{fig:basic_characterization}
\end{figure}

We designed an integer fluxonium for erasure conversion and detection, incorporating its readout and control lines as illustrated in \fref{fig:basic_characterization}(a). The integer fluxonium is coupled to a charge line and flux line for microwave drives, and a readout resonator for state measurement. We stress that both the mid-circuit erasure checks and the final computational state readout are performed using the same coupled resonator, reducing the hardware overhead. The device was fabricated and packaged at MIT Lincoln Laboratory, following the process in \cite{Gingras2026_MITLLFab}.

With the prepared device, we performed single-tone and two-tone spectroscopy to extract the readout and qubit parameters. Fitting the measured spectrum yielded fluxonium parameters of $E_{C}/h=$\SI{1.392}{GHz}, $E_{J}/h=$\SI{5.056}{GHz}, $E_{L}/h=$\SI{0.193}{GHz}, confirming that the device operates within the integer fluxonium regime. The qubit spectrum near zero flux bias is shown in \fref{fig:basic_characterization}(b). At the zero-flux sweet spot, we measured a qubit g-e transition frequency of \SI{3.5496}{GHz} and e-f transition frequency of \SI{30.53}{MHz}. Notably, the resonant frequency shift between the ground state and the erasure state ($\chi_{\mathrm{ge}}/2\pi$) was 1.86$\pm$\SI{0.03}{MHz}, while the resonant frequency shift between the two logical states ($\chi_{\mathrm{gf}}/2\pi$) was suppressed to -11.9$\pm$\SI{13.8}{kHz} [\fref{fig:basic_characterization}(d)]. This result satisfies our design goal of $|\chi_{\mathrm{ge}}| \gg |\chi_{\mathrm{gf}}|$, which enables ancilla-free erasure detection by minimizing the dephasing within the computational states.

Next, we characterized the energy relaxation time for the $\mathrm{f \rightarrow e}$ and $\mathrm{e \rightarrow g}$ transitions at zero flux. The qubit was initialized in $|\mathrm{f} \rangle$ using two sequential g-e and e-f $\pi$-pulses, after which we measured the population evolution of the $\mathrm{|g\rangle,|e\rangle}$, and $\mathrm{|f \rangle}$ states. The resulting data were fit using a three-state decay model as detailed in \aref{app:decay_numerical_fit}. From this numerical fit, we obtained $T_{1}^{\mathrm{f \rightarrow e}} \approx~$400 -- \SI{600}{\micro s}, $T_{1}^{\mathrm{e \rightarrow g}} \approx~$200 -- \SI{300}{\micro s} [\fref{fig:basic_characterization}(c)]. The direct relaxation time from $\mathrm{|f\rangle}$ to $\mathrm{|g\rangle}$ was estimated at $T_{1}^{\mathrm{f \rightarrow g}}=$ 4 -- \SI{6}{ms} from the fit, confirming that the majority of the decay events from the logical state $|\mathrm{f} \rangle$ pass through the detectable erasure state $|\mathrm{e} \rangle$.

\begin{figure*}
    \includegraphics[width=0.99\linewidth]{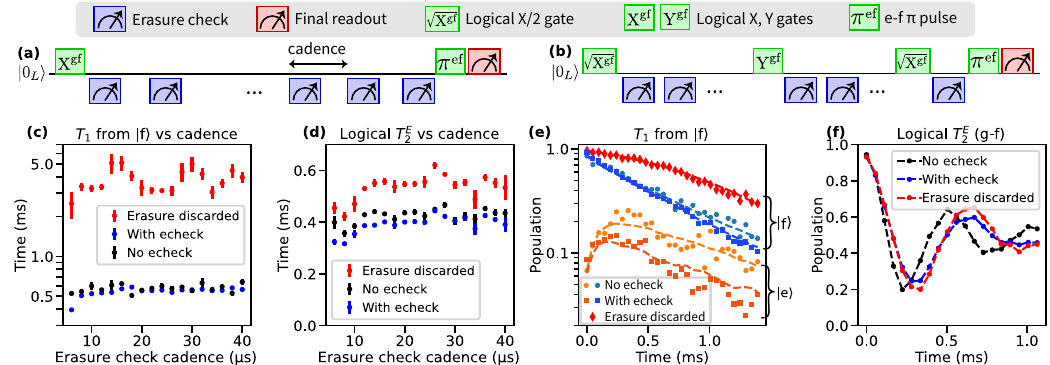}
    \caption{(a, b) Pulse sequences for measuring (a) $T_{1}$ and (b) $T_{2}^{E}$ including mid-circuit erasure checks and final readout. (c) $T_{1}$ from $|\mathrm{f} \rangle $ for three cases: without erasure checks, with erasure checks but no postselection, and after discarding erasure error events. The postselected $T_{1}$ reflects the $T_{1}^{\mathrm{f \rightarrow g}}$ extracted from the numerical fit. (d) $T_{2}^{E}$ within computational states $|\mathrm{g}\rangle, |\mathrm{f}\rangle$. As in panel (c), the data include without erasure checks, with erasure checks, and after discarding erasure errors. (e) Population evolution of the $|\mathrm{f}\rangle$ and $|\mathrm{e} \rangle$ states initialized in $|\mathrm{f}\rangle$, measured without erasure checks and with mid-circuit erasure checks at \SI{16}{\micro s} cadence. Dashed lines represent numerical fits to the experimental data (markers). We observed a significant extension of the logical lifetime after discarding the erasure errors. (f) Population evolution of a Hahn-echo sequence at the erasure check cadence of \SI{26}{\micro s}. We measured 38\% increase of $T_{2}^{E}$ after discarding the erasure error. Error bars represent standard error across five repeated measurements and characterize run-to-run variation in the measured coherence times.}
    \label{fig:echeck_with_decoherence}
\end{figure*}

\subsection{Measurements of coherence times}
Following the basic characterization, we implemented mid-circuit erasure checks to verify the erasure-check capability. We evaluated the effectiveness of this approach by comparing the qubit coherence times under three conditions: without erasure checks, with erasure checks, and after postselecting the data against the erasure error events.

The erasure check is performed by applying a microwave drive to the readout resonator near its resonant frequency when the qubit is in the erasure state $|\mathrm{e} \rangle$. The amplitude, duration, and frequency of the erasure check were chosen to have a high-fidelity discrimination between the erasure state and the computational states in the IQ plane. Simultaneously, the parameters were constrained to suppress the erasure-check-induced dephasing within the logical states, as $\chi_{\mathrm{gf}}$ remains small but finite. For these measurements, the erasure check frequency was set to \SI{7.39020}{GHz}, which is detuned by \SI{2.18}{MHz} from the logical state resonant frequency \SI{7.39238}{GHz}, and \SI{0.32}{MHz} detuned from the erasure state resonance \SI{7.39052}{GHz}. The erasure check duration and amplitude were \SI{1.1}{\micro s} and the average photon number of $\bar{n}\approx 1.7$ when the state is in $|\mathrm{e}\rangle$. 

After setting the erasure check parameters, we investigated the impact of the erasure check on the $|\mathrm{f} \rangle$ state lifetime as we varied the duration between consecutive erasure checks (cadence). The lifetime of the $|\mathrm{f} \rangle$ state without removing the erasure error was estimated by fitting the decay curve to the exponential decay equation $Ae^{-t/T_{1}}+ B$. After removing the shots in which an erasure error was detected, we observed the improvement in the $|f\rangle$ state lifetime [\fref{fig:echeck_with_decoherence}(e)]. Notably, the postselected decay curve exhibited non-exponential behavior (\fref{fig:echeck_with_decoherence}(e)) due to an increased incidence of missed erasure events ($\mathrm{|f\rangle \rightarrow |e\rangle \rightarrow |g\rangle}$). To account for this, we estimated the lifetime in the short-time limit, where dynamics are dominated by the $T_{1}^{\mathrm{f \rightarrow }g}$ relaxation. We achieved a peak $|\mathrm{f} \rangle$ state lifetime of 5.087$\pm$\SI{0.685}{ms} at an erasure check cadence of \SI{16}{\micro s} after postselecting against the erasure error events, which is an 8.4-fold increase over the lifetime without erasure checks, 0.606$\pm$\SI{0.054}{ms}.

Next, we measured the Hahn-echo time ($T_{2}^{E}$) within the computational subspace, comparing performance with and without erasure checks [\fref{fig:echeck_with_decoherence}(d, f)]. Baseline measurements without erasure checks resulted in $T_{2}^{E}$ between 400--\SI{500}{\micro s}, representing the variation across the sampled erasure check cadences. After removing the erasure error, $T_{2}^{E}$ improved to 550--\SI{600}{\micro s}, with the maximum value of 620$\pm$\SI{14}{\micro s} at \SI{26}{\micro s} cadence. This constitutes a 38\% increase from the baseline $T_{2}^{E}$ of 450$\pm$\SI{10}{\micro s} without erasure checks. 

In summary, we have successfully demonstrated the conditional increase in both $T_{1}$ and $T_{2}^{E}$ after discarding erasure errors detected by mid-circuit erasure checks. These results validate the erasure-conversion capability of an integer fluxonium.


\begin{figure*}[t]
    \centering
    \includegraphics[width=0.95\linewidth]{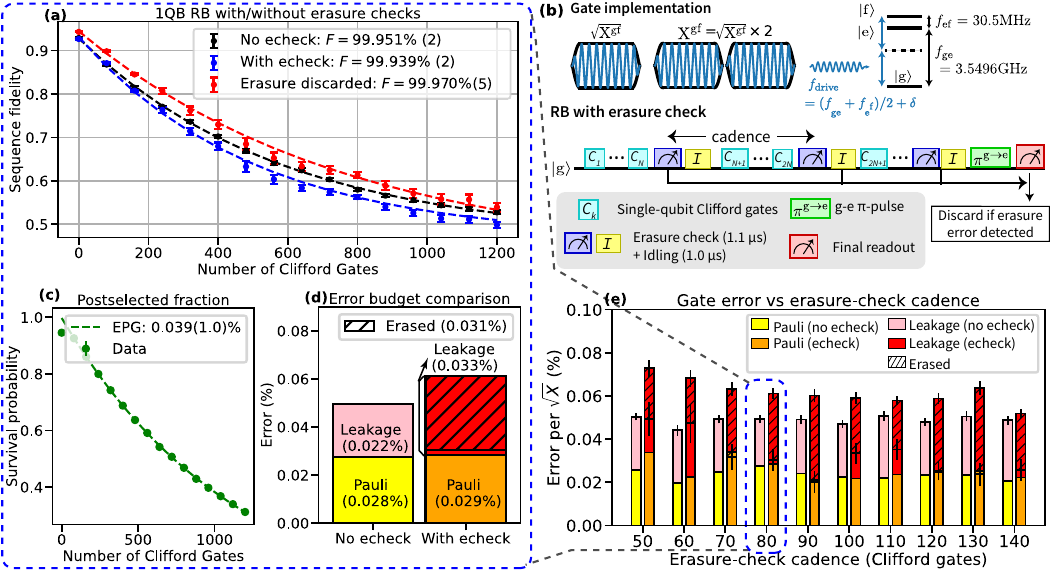}
    \caption{(a) RB results without erasure checks (black), with erasure checks (blue), and after discarding erasure errors (red). The plot shows the result when the erasure-check cadence of 80 Clifford gates. $F$ denotes the average gate fidelities per $\sqrt{X}$, which are obtained from LRB analysis. (b) Schematics of gate implementation and the RB pulse sequence. The $\sqrt{X}$, $X$ gates were implemented via microwave pulses with a frequency near the half of the g-f transition of the qubit. The RB sequence consists of random single-qubit Clifford gates $C_{k}$ with the mid-circuit erasure checks interleaved periodically. (c) Postselected fraction by the number of Clifford gates when mid-circuit erasure checks are present. An exponential fit (green dashed line) yields a postselection rate of 0.039(1)\% per $\sqrt{X}$. (d) Error rate breakdown for the qubit without (left) and with (right) the erasure checks. Colored bars represent the Pauli error (yellow, orange), leakage (pink, red), and discarded leakage components (hatched). We discarded 94\% of the leakage error via the mid-circuit erasure check results. (e) Error per $\sqrt{X}$ measured by the erasure-check cadence. The amount of leakage error showed an increasing trend for more frequent erasure checks (decreasing cadence), while the Pauli errors were maintained at similar level.}
    \label{fig:1qbrb_with_echeck}
\end{figure*}

\subsection{\label{sec:gate_calibration} Single-qubit gate calibration}

After measuring the coherence time improvement after postselecting against the erasure error events, we implemented and calibrated a single-qubit gate. The single-qubit gates were implemented using a microwave drive with a pulse envelope consisting of \SI{10}{ns} rise and fall times and a \SI{70}{ns} plateau with an additional \SI{10}{ns} buffer between pulses. We employed a square-root cosine envelope for the rise and fall; since the g-f transition is a two-photon process, this results in an effective cosine-shaped turn-on.

The gate amplitude and frequency were calibrated using error-amplifying pulse sequences. We utilized odd numbers of $X$ gates to calibrate overrotations (amplitude), and sequences of alternating pairs of $X$ gates $[X,-X]$ to amplify frequency and phase errors. Due to the significant AC Stark shift inherent in the high-power two-photon drive, the gate amplitude and frequency cannot be calibrated independently. The detailed procedure of gate calibration is shown in \fref{fig:1qb_gate_calibration} in \aref{app:gate_calibration_procedure}. 

\subsection{\label{sec:gate_benchmarking} Single-qubit gate benchmarking}

Following single-qubit gate calibration, we performed Clifford randomized benchmarking (RB)~\cite{Emerson2005_CliffordRB} across 20 different randomized sequences and compared the results with and without erasure checks. With erasure checks, the checks were interleaved periodically from 50 Clifford gates to 140 Clifford gates. The resulting data were processed using the leakage RB (LRB) analysis, which extracts the leakage rate and gate fidelity by fitting the sequence fidelity and the computational-subspace probability. Detailed procedures of the LRB analysis are provided in \aref{app:lrb_protocol} and~\cite{Wood2018_LRB, RuiLi2024_LRBexample, RuiLi2025_lrb_example}. When erasure checks were not included in RB sequence, the computational-subspace population was measured separately by removing the g-e $\pi$-pulse before the final readout, which can distinguish shots in $|\mathrm{e}\rangle$ from $\mathrm{|g\rangle, |f\rangle}$.

For each erasure-check cadence, the leakage error was obtained from LRB analysis, and subsequently the Pauli error was estimated by subtracting the leakage error from the total error. Finally, the amount of detected leakage was estimated from mid-circuit erasure checks, representing the effectiveness of erasure conversion. 

In \fref{fig:1qbrb_with_echeck}(a, c, d), results from erasure-check cadence of 80 Clifford gates are presented. From the LRB analysis, we obtained a baseline average gate fidelity of 99.951(2)\% without erasure checks. Upon introducing the erasure checks but not postselecting, the average gate fidelity decreased to 99.939(2)\% [\fref{fig:1qbrb_with_echeck}(a)]. Using the gate fidelity and leakage error rate obtained, we constructed a detailed error budget of the system [\fref{fig:1qbrb_with_echeck}(d)]. In the absence of erasure checks, we estimate a leakage error rate of 0.022(2)\%, with the remaining error (0.028\%) attributed to Pauli errors. With erasure checks included, the leakage error rate increased to 0.033(6)\%. This increment in leakage (0.011\%) matches the baseline fidelity drop when the erasure check was added. For different erasure-check cadence, we consistently observed increase in leakage errors compared to when erasure check was not present, suggesting the addition of erasure-checks largely adds the leakage errors instead of Pauli errors.

\begin{figure*}
    \centering
    \includegraphics[width=0.99\linewidth]{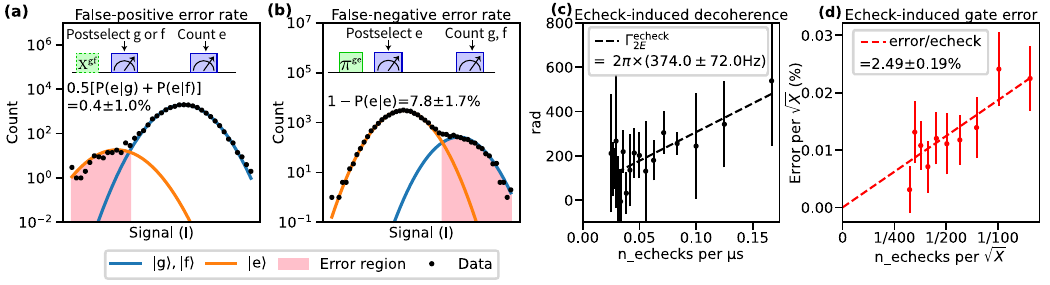}
    \caption{(a, b) Histograms of the integrated readout signal $I$ used to determine the (a) false-positive and (b) false-negative error rates of the mid-circuit erasure check. Black markers represent experimental data and solid lines denote Gaussian fits. The shaded pink regions indicate the classification error region. The data were acquired using two readouts: first erasure check to postselect the initial state, followed by a subsequent erasure check to count classification errors. We measured a false-positive error rate of 0.4$\pm$1.0\%, and false-negative error rate of 7.8$\pm$1.7\%. (c) Erasure-check-induced decoherence rate as a function of erasure-check cadence. The erasure-check-induced decoherence rate (black) was obtained by subtracting the baseline Hahn echo rate (without erasure check) from the rates measured with mid-circuit erasure checks [\fref{fig:echeck_with_decoherence}(d)]. A linear fit extracts an induced decoherence rate of 374.0$\pm$\SI{72.0}{Hz}. (d) Erasure-check-induced gate error as a function of number of erasure checks per $\sqrt{X}$ gates. The data was obtained by subtracting the gate error without erasure check from when the erasure checks were present [\fref{fig:1qbrb_with_echeck}(e)]. The data were fit using the linear function with y-intercept forced to zero. We obtained error per erasure check of 2.49$\pm$0.19\% from the slope of the linear fit.} 
    \label{fig:echeck_analysis}
\end{figure*}

To validate the consistency of the LRB analysis, we analyzed the postselected fraction as a function of the number of Clifford gates [\fref{fig:1qbrb_with_echeck}(c)]. An exponential fit yielded an erasure error rate of 0.039(1)\% per $\sqrt{X}$ gate, which is in agreement with the leakage error rate 0.033(6)\% extracted via LRB.

By performing mid-circuit erasure checks, we identified 94\% of the leakage errors. This postselection increased the average gate fidelity to 99.970(5)\%, at which the system becomes limited primarily by the residual Pauli errors. This result constitutes a clear demonstration of the efficacy of the erasure checks in an integer fluxonium architecture. 

In summary, our randomized benchmarking results demonstrate that mid-circuit erasure checks provide a considerable performance gain. By identifying that nearly half of the total gate error is due to detectable leakage and successfully discarding over $90\%$ of these events, we achieved a fidelity improvement from $99.939(2)\%$ to $99.970(5)\%$. This represents a $50\%$ reduction in the average error per gate, proving that the integer fluxonium architecture can enhance quantum gate operations through erasure checks. However, the current device does not satisfy the erasure-bias condition as the remaining Pauli errors were comparable to the leakage error.

\subsection{\label{subsec:error_from_echcks} Error from the erasure checks}

We characterized the performance of each erasure check by calculating four figures of merits: false-positive error rate, false-negative error rate, erasure-check-induced decoherence, and erasure-check-induced gate error. 

\subsubsection{False-positive error rate}
The false-positive error rate denotes the probability that a logical state ($\mathrm{|f\rangle}$ or $\mathrm{|g \rangle}$) is misclassified as the erasure state $|\mathrm{e}\rangle$. We calculated this error rate as $\frac{1}{2} \left[\mathrm{P(e|f) + P(e|g)}\right]$, for which we measured a value of 0.4$\pm$1.0\% [\fref{fig:echeck_analysis}(a)]. This error predominantly occurred when the system was initialized in $|\mathrm{f} \rangle$, which can be seen in \tref{tab:assignment_fidelities}. 

\subsubsection{False-negative error rate}

False-negative error rate is the probability of failing to detect a prepared erasure state, calculated as $1 - \mathrm{P(e|e)}$. We measured a significantly higher false-negative error rate of 7.8$\pm$1.7\% compared to false-positive error rate. This value exceeds the limit expected from intrinsic relaxation alone. We attribute this enhancement to a reduction in $T_{1}^{\mathrm{e \rightarrow g}}$ relaxation time during the erasure check, which dropped from several hundred microseconds to \SI{20}{\micro s} while the erasure check was active. This measurement is detailed in \aref{app:T1_reduction_during_echeck}. Given the erasure check duration of \SI{1.1}{\micro s}, we estimate the decay probability during the check to be $1.1/20 \approx~$5.5\%, which is the primary contributor to the observed false-negative error rate. 

\subsubsection{Erasure-check-induced decoherence}

The erasure-check-induced decoherence was calculated by comparing the Hahn-echo decoherence rates measured with and without the erasure checks across various cadences. From a linear fit, we extracted the erasure-check-induced decoherence rate of $\Delta \Gamma_{2E}^{\mathrm{echeck}}/2\pi=$374$\pm$\SI{72}{Hz} [\fref{fig:echeck_analysis}(c)]. 

We compared the erasure-check-induced decoherence rate with the theoretical model of measurement-induced dephasing rate. From the logical state resonant frequency shift and the average photon number, we estimate the erasure-check-induced dephasing as follows~\cite{Gambetta2006_memt_induced_dephasing}.
\begin{align}
    \Gamma_{\varphi }^{\mathrm{eras}} = \frac{(\bar{n}_{\mathrm{g}} + \bar{n}_{\mathrm{f}})\kappa (\chi_{\mathrm{gf}}/2)^{2}}{(\kappa/2)^{2} + (\chi_{\mathrm{gf}}/2)^{2} + (2\pi f_{\mathrm{res}}^{0}-2\pi f_{d})^{2}}
\end{align}
where $f_{\mathrm{res}}^{0}$ is the bare resonator frequency and $f_{d}$ is the erasure-check frequency. $\bar{n}_{\mathrm{g}}$, $\bar{n}_{\mathrm{f}}$ are the average photon numbers when the state is in $|\mathrm{g}\rangle$, $|\mathrm{f} \rangle$, respectively, satisfy the following equation.
\begin{align}
    \bar{n}_{\mathrm{g,f}} = \frac{\varepsilon_{\mathrm{rf}}^{2}}{(\kappa/2)^{2} + (2\pi f_{d} - 2\pi f_{\mathrm{res}}^{\mathrm{|g\rangle, |f\rangle}})^{2}}
\end{align}
where $\varepsilon_{\mathrm{rf}}$ is drive amplitude proportional to the readout input signal level. In this setup, we estimate $\varepsilon_{\mathrm{rf}} \approx 7.0 \times 10^{6}$ and the other parameter values are listed in \tref{tab:readout_parameters} of \aref{app:system_parameters}. Under these parameters, we expect $\bar{n}_{\mathrm{g,f}}\approx 0.22$ and the erasure-check-induced dephasing rate to be $\Gamma_{\mathrm{eras}}/2\pi=$\SI{17.6}{Hz}. 

The discrepancy between the measured (\SI{374}{Hz}) and predicted (\SI{17.6}{Hz}) induced error rates is consistent with most of the excess decoherence induced by the erasure check does not originate primarily from pure dephasing, but rather from an enhanced $T_{1}$ relaxation of the logical states. For instance, the erasure check may facilitate the cascaded relaxation process $\mathrm{|f\rangle \rightarrow |e\rangle \rightarrow |g\rangle}$ by decreasing the relaxation time from $|\mathrm{e}\rangle$ to $|\mathrm{g} \rangle$, as detailed in \aref{app:T1_reduction_during_echeck}.   

\subsubsection{Erasure-check-induced gate error}

Lastly, the erasure-check-induced gate error is obtained by comparing the average gate fidelity with and without the erasure checks as a function of number of erasure checks per $\sqrt{X}$ gate [\fref{fig:echeck_analysis}(d)]. From the linear fit, we estimate each erasure check adds 2.49$\pm$0.19\% additional error. Since each erasure check adds additional time of \SI{2.1}{\micro s}, the incoherent error added for each erasure check is going to be~\cite{Abad2024_gatefidelity}:
\begin{align}
\begin{split}
    r_{\mathrm{incoh}} & \approx  \frac{2.1\mathrm{\mu s}}{2} \frac{1}{T_{1}^{\mathrm{f \rightarrow e}}} + \frac{2.1 \mu s}{3}  \frac{1}{T_{\varphi E}} + \frac{2(1.1 \mathrm{\mu} s)}{3} \Delta \Gamma_{2E}^{\mathrm{echeck}} \\ 
    & \approx \frac{2.1 \mathrm{\mu s}}{3} \frac{1}{T_{1}^{\mathrm{f \rightarrow e}}} + \frac{2.1 \mu s}{3}  \frac{1}{T_{2 E}} + \frac{2(1.1 \mu s)}{3} \Delta \Gamma_{2E}^{\mathrm{echeck}} \\
    & \approx 0.46\%    
\end{split}
\end{align}
where we substituted $T_{1}^{\mathrm{f \rightarrow e}} \approx~$\SI{600}{\micro s} and $T_{2E} \approx~$\SI{400}{\micro s}, and $\Delta \Gamma_{2E}^{\mathrm{echeck}}=2\pi \times 374 \mathrm{Hz}$, obtained from \fref{fig:echeck_analysis}(c). The coefficient $2/3$ multiplied to $\Delta \Gamma_{2E}$ is from the assumption that most of the erasure-check-induced decoherence stems from the logical relaxation $\mathrm{|f\rangle \rightarrow |g\rangle}$ and not from the pure dephasing. The additional incoherent error during each erasure check only constitutes 18\% of the total error induced by erasure check. The source of additional error is currently unknown, but it is possible that each erasure check may induce a large amount of error mediated by transition to the non-computational states beyond the three-level system of the g-f erasure qubit. 

In summary, our mid-circuit erasure check demonstrates a low false-positive rate. However, the false-negative rate remains high due to measurement-induced $\mathrm{|e\rangle \rightarrow |g \rangle}$ transition, which might be mitigated in future iterations through dedicated ancilla qubits for readout or targeted engineering of the two-level system bath to stabilize the $|\mathrm{e}\rangle$ state. Furthermore, we measured a 2.49\% induced error per erasure check, which cannot be explained by the estimated incoherent error budget. Identifying the origin of this excess error remains an important goal for future work.

\section{Conclusion and Discussion}

In conclusion, we have demonstrated erasure conversions and direct mid-circuit erasure detection in a single integer fluxonium. We achieved 8.4-fold improvement of $T_{1}$ from 606$\pm$\SI{54}{\micro s} to 5.087$\pm$\SI{0.685}{ms} and 38\% increase of $T_{2}^{E}$ from 450$\pm$\SI{10}{\micro s} to 620$\pm$\SI{14}{\micro s} after postselecting against detected erasures. In addition, we achieved a single-qubit gate fidelity improvement from 99.939(2)\% to 99.970(5)\% after discarding the erasures detected during mid-circuit erasure checks. Crucially, by performing erasure checks directly through the coupled readout resonator, we demonstrated erasure detection and total error reduction via postselection is achievable with a reduced hardware footprint.

The current device, however, faces specific challenges to overcome before it can qualify as an effective erasure qubit. The LRB results suggest that the residual Pauli errors remain comparable to the leakage error, which does not fully satisfy the erasure-bias condition. To realize a strongly erasure-biased noise model, the Pauli errors must be reduced, potentially through careful pulse optimization or flux stabilization. 

In addition, we observed a significant increase in the $\mathrm{|e \rangle \rightarrow |g \rangle }$ decay rate when the erasure check signal is active, which increases the probability of missed erasure errors. Missed erasure errors can be converted into an undetected Pauli error, thereby degrading the overall erasure bias. While this effect may be partially mitigated by optimizing the drive frequency and amplitude of the erasure check, it remains a major challenge for the present ancilla-free architecture. Future iterations may require introducing an ancilla qubit for the erasure check, albeit at the cost of increased hardware complexity.

Beyond hardware efficiency, the integer fluxonium offers additional practical advantages. First, by operating at the zero-flux sweet spot, the requisite current bias can be considerably smaller than fluxonium biased at half a flux quantum, thereby reducing the potential heating from the bias lines. Furthermore, the erasure channel is strongly state asymmetric: erasure events occur predominantly from the $|\mathrm{f}\rangle$ state rather than the $|\mathrm{g}\rangle$ state. This asymmetric error distribution provides additional prior information that can be leveraged to further enhance the efficiency and accuracy of erasure-aware decoders~\cite{Sahay2023_eraserror_benefit_neutralatom}. 

Overall, this work provides the demonstration of an ancilla-free erasure check within a single-qubit footprint. Ultimately, the integer fluxonium offers a feasible solution to bridge the gap between hardware simplicity and robust erasure error detection.\\ 

\noindent \textit{Note added}.--We recently became aware of closely related work by Liu \textit{et al}.~\cite{ChenluLiu2026_gfflxnm}.

\section{\label{sec:acknowlegements} ACKNOWLEDGMENTS}
We gratefully acknowledge fruitful conversation with~Xanthe Croot,~Stefan Filipp, Longxiang~Huang, Eli Levenson-Falk,~Klaus Liegener,~Xizheng Ma,~Johannes Schirk,~Christian Schneider,~Sam Taubenberger, and~Florian Wallner. This research is sponsored in part by the US Army Research Office grant number W911NFF-23-1-0045 (Extensible and Modular Advanced Qubits); 
in part by the US Department of Energy, Office of Science, National Quantum Information Science Research Centers, Co-design Center for Quantum Advantage (C2QA), under contract number DE-SC0012704;
and in part under Air Force contract number FA8702-15-D-0001.
J.A. and J.K. gratefully acknowledge support from Korea Foundation for Advanced Studies. 
M.H. is supported by an appointment to the Intelligence Community Postdoctoral Research Fellowship Program at the Massachusetts Institute of Technology administered by Oak Ridge Institute for Science and Education (ORISE) through an interagency agreement between the U.S. Department of Energy and the Office of the Director of National Intelligence (ODNI).
\\

The views and conclusions contained herein are those of the authors and should not be interpreted as necessarily representing the official policies or endorsements, either expressed or implied, of the U.S. Government.
\clearpage 

\providecommand{\noopsort}[1]{}\providecommand{\singleletter}[1]{#1}%

\clearpage
\newpage 
\appendix

\section{MEASUREMENT SETUP}
Measurements of the integer fluxonium were conducted in a Bluefors XLD600 dilution refrigerator, operating at a base temperature near \SI{20}{mK}. The chip was affixed to a cold finger on the mixing chamber stage and shielded by a Mu-metal enclosure.

The readout signal was created by a Holzworth HS9001B RF synthesizer and combined with a signal from a Keysight M3202 PXIe arbitrary waveform generator (AWG) through a mixer. The generated readout signal was attenuated by \SI{70}{dB} and filtered by a \SI{12}{GHz} low-pass filter, and an Eccosorb filter to suppress thermal noise and infrared radiation. Following reflection from the chip, the signal underwent initial amplification via a Josephson traveling-wave parametric amplifier (JTWPA) on the mixing chamber stage. The JTWPA was activated by a pump signal from a Holzworth HS9001B RF synthesizer, which was attenuated by \SI{60}{dB} prior to reaching the JTWPA. The JTWPA’s output signal was subsequently amplified by two high-electron-mobility-transistor amplifiers, one located at the \SI{4}{K} stage and the other at room temperature. Following this, the signal was down-converted by a mixer, then amplified with a Stanford Research SR445A preamplifier, and finally demodulated using a Keysight M3102A Analog-to-Digital Converter (ADC or digitizer).

The microwave drive signal that drives the g-e transition was first generated by a R\&S SGS100A SGMA RF signal generator, and pulse-modulated by a Keysight M3202 AWG signal through an internal IQ mixer of R\&S SGS100A. The microwave drive signal was attenuated by \SI{60}{dB} and filtered by a low-pass filter and an eccosorb to reduce the thermal noise and infrared radiation. The microwave drive signal that drives e-f transition was generated by a Keysight M3202 AWG signal alone because the transition frequency is within the bandwidth of the AWG.  

The DC current for flux control was generated by a Yokogawa GS200 DC source. Two 2000 ohm resistors were placed in series with the two DC-source leads to attenuate the noise from the instrument. The DC current was eventually applied to the global coil mounted on top of the package that was used to control the magnetic field applied to the qubit.

\begin{figure}[H]
    \centering
    \includegraphics[width=0.99\linewidth]{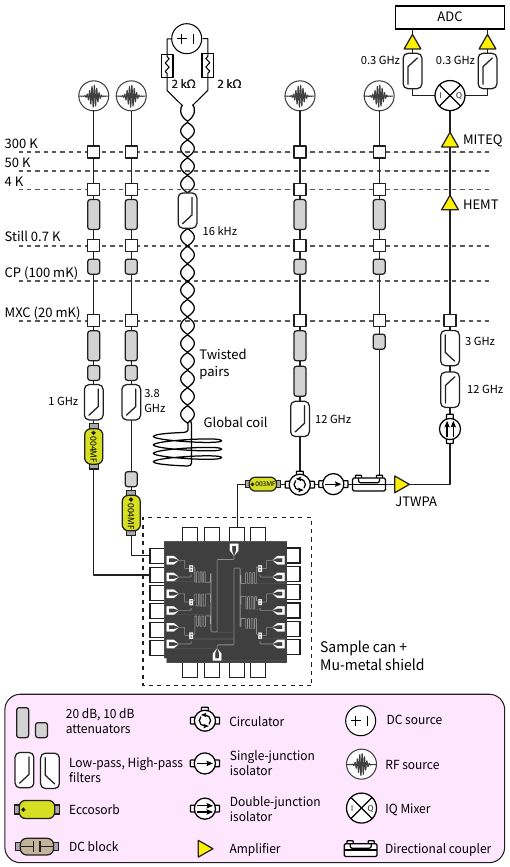}
    \caption{Room-temperature and cryogenic wiring of the experimental setup.}
    \label{fig:wiring_diagram}
\end{figure}

We refer to \fref{fig:wiring_diagram} for a more detailed layout of the wiring. Also, the room-temperature electronics are listed in table \ref{tab:instruments}.

\begin{table}[H]
    \centering
    \begin{tabular}{c|c|c} 
        \hline 
        \hline
        Instrument type & Manufacturer & Model \\
        \hline 
        DC source (global coil) & Yokogawa & GS200 \\
        RF source & Rohde \& Schwarz & SGS100A \\
        RF source & Holzworth & HS9001B \\ 
        Preamplifier & Stanford Research & SR445A \\ 
        Control chassis & Keysight & M9019A \\
        AWG & Keysight & M3202A \\ 
        ADC & Keysight & M3102A \\
         \hline
         \hline
    \end{tabular}
    \caption{Room-temperature electronic instruments used in this research.}
    \label{tab:instruments}
\end{table}

\clearpage

\onecolumngrid
\begin{widetext}

\section{SYSTEM PARAMETERS} \label{app:system_parameters}

\begin{table}[H]
    \centering
    \begin{tabular}{l|c|c}
        \hline
        \hline
        Readout parameters & Cycle 1 & Cycle 2 \\ 
        \hline
        Bare resonator frequency($f_{\mathrm{res}}^{0}$) & \SI{7.39109}{GHz} & \SI{7.39009}{GHz} \\
        Erasure check frequency & \SI{7.3902}{GHz} & \SI{7.38964}{GHz}\\
       Resonant frequency at $\mathrm{|g\rangle}$($f_{\mathrm{res}}^{|\mathrm{g} \rangle}$)& \SI{7.39242}{GHz} & \SI{7.39139}{GHz}\\
       Resonant frequency at $|\mathrm{e} \rangle$($f_{\mathrm{res}}^{|\mathrm{e} \rangle}$)& \SI{7.39052}{GHz} & \SI{7.38960}{GHz} \\
        $f_{\mathrm{res}}^{|\mathrm{g} \rangle} - f_{\mathrm{res}}^{|\mathrm{e} \rangle}$ ($\chi_{\mathrm{ge}}/2\pi$)  & {1.86}$\pm$\SI{0.03}{MHz} & 1.79$\pm$\SI{0.01}{MHz}\\ 
        $f_{\mathrm{res}}^{|\mathrm{g} \rangle} - f_{\mathrm{res}}^{|\mathrm{f} \rangle}$ ($\chi_{\mathrm{gf}}/2\pi$) & -11.9$\pm$\SI{13.8}{kHz} & -27.9$\pm$\SI{14.0}{kHz}\\
        Resonator \\ internal coupling ($\kappa_{i}/2\pi$)& \SI{0.078}{MHz} & \SI{0.439}{MHz} \\
        Resonator \\ external coupling ($\kappa_{c}/2\pi$)& \SI{1.370}{MHz} & \SI{1.144}{MHz} \\
        \hline
        \hline
        Fluxonium qubit parameters & Cycle 1 & Cycle 2 \\ 
        \hline
        $E_{C}/h$ & \SI{1.392}{GHz} & \SI{1.398}{GHz} \\
        $E_{J}/h$ & \SI{5.056}{GHz} & \SI{5.088}{GHz} \\
        $E_{L}/h$ & \SI{0.193}{GHz} & \SI{0.193}{GHz} \\ 
        Qubit-resonator coupling ($g/2\pi$) \\ ($\hat{H} = \hbar g \hat{n}_{\mathrm{qb}} (\hat{a}+ \hat{a}^{\dagger})$) & \SI{97}{MHz} & \SI{96.4}{MHz} \\
        \hline 
        \hline
        Coherence Time & Cycle 1 & Cycle 2 \\ 
        \hline
        $T_{2R}^{\mathrm{gf}} $ (without echeck) & -- & 357$\pm$\SI{76}{\micro s} \\
        $T_{2R}^{\mathrm{gf}} $ (\SI{10}{\micro s} echeck cadence) & -- & 317$\pm$\SI{29}{\micro s} \\
        
    \end{tabular}
    \caption{Readout and qubit parameters of the integer fluxonium measured in this paper. cycle 1 and Cycle 2 refer to the two different measurement sessions performed at different cooldown.}
    \label{tab:readout_parameters}
\end{table}
\end{widetext}
\section{THREE-STATE READOUT IN AN INTEGER FLUXONIUM}
\begin{figure}[H]
    \centering
    \includegraphics[width=0.99\linewidth]{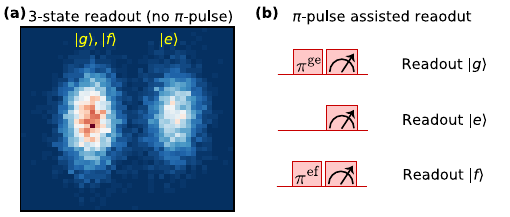}
    \caption{(a) IQ plane histogram of the single-shot distributions of the three states $\mathrm{|g\rangle, |e\rangle, |f\rangle}$. $\mathrm{|g\rangle, |f\rangle}$ are indistinguishable, which is reasonable since $\chi_{\mathrm{gf}}$ is small. (b) Pulse configuration of $\pi$-pulse assisted readout. By repeating the experiment with different pre-readout mapping pulses, we can separately estimate the population of all three states.}
    \label{fig:threestate_ro}
\end{figure}

Because the $|\mathrm{g} \rangle$ and $|\mathrm{f} \rangle$ states were indistinguishable in this study (\fref{fig:threestate_ro}(a)), we applied a $\pi$-pulse before the readout to distinguish the two computational states [\fref{fig:threestate_ro}(b)]. To detect the $|\mathrm{g} \rangle $ state, we applied a $\pi$-pulse from $|\mathrm{g}\rangle$ to $|\mathrm{e} \rangle$ ($\pi^{\mathrm{g-e}}$) before the final readout. Similarly, a $\pi$-pulse between $|\mathrm{e}\rangle$ and $|\mathrm{f} \rangle$ ($\pi^{\mathrm{e-f}}$) was applied before the final readout to detect the $|\mathrm{f} \rangle$ state. The following table presents the assignment fidelity across three states.

\begin{table}[H]
    \centering
    \begin{tabular}{c|c|c|c}
    \hline
    \hline
   Final readout & Prepare $|\mathrm{g} \rangle$ & Prepare $|\mathrm{e} \rangle $ & Prepare $|\mathrm{f} \rangle $ \\ 
    \hline
        Measure $\mathrm{|g}\rangle$ & 97.4\% & 3.5\% & 1.0\%\\
        Measure $\mathrm{|e}\rangle$ & 0.3\% & 94.8\% & 0.8\%\\ 
        Measure $\mathrm{|f}\rangle$ & 0.3\% & 2.6\% & 95.2\%\\ 
    \hline 
    \hline
    \end{tabular}
    \caption{State preparation and assignment fidelity for the final readout. Because the three populations are inferred from separate mapping-pulse experiments rather than from a single readout, the entries in each preparation column are not constrained to sum exactly to 100\%.}
    \label{tab:assignment_fidelities}
\end{table}


\clearpage
\section{\label{app:decay_numerical_fit} NUMERICAL SIMULATION OF DECAY WITHIN g-f ERASURE QUBIT}
In this section, we describe the two numerical methods used to simulate the population evolution of the g-f erasure qubit. While our analysis focuses on the integer fluxonium architecture, these methods are applicable to three-level systems whose dynamics are dominated by Markovian incoherent population transitions. 

\subsection{Three-state decay model}

The first method solves a system of ordinary differential equations (ODEs) that describe the time-evolution of the occupancy probabilities ($p_{\mathrm{g}}, p_{\mathrm{e}}, p_{\mathrm{f}}$) for $\mathrm{|g\rangle,|e\rangle,|f\rangle}$ states, respectively: 

\begin{align}\label{eqn:three_state_decay_ode}
\begin{split}
    \frac{{dp}_{\mathrm{f}}}{dt} &= -(\Gamma_{1}^{\mathrm{f \rightarrow e}} + \Gamma_{1}^{\mathrm{f \rightarrow g}}) p_{\mathrm{f}} + \Gamma_{1}^{\mathrm{e \rightarrow f}} p_{\mathrm{e}} + \Gamma_{1}^{\mathrm{g \rightarrow f}} p_{\mathrm{g}} \\
    \frac{d p_{\mathrm{e}}}{dt} &=\Gamma_{1}^{\mathrm{f \rightarrow e}} p_{\mathrm{f}} - (\Gamma_{1}^{\mathrm{e \rightarrow f}} + \Gamma_{1}^{\mathrm{e \rightarrow g}} )p_{\mathrm{e}} + \Gamma_{1}^{\mathrm{g \rightarrow e}} p_{\mathrm{g}} \\
    \frac{d p_{\mathrm{g}}}{dt} &=\Gamma_{1}^{\mathrm{f \rightarrow g}} p_{\mathrm{f}} + \Gamma_{1}^{\mathrm{e \rightarrow g}} p_{\mathrm{e}} -( \Gamma_{1}^{\mathrm{g \rightarrow e}} + \Gamma_{1}^{\mathrm{g \rightarrow f}}) p_{\mathrm{g}}.
\end{split}
\end{align}

\eref{eqn:three_state_decay_ode} can be solved numerically using standard numerical libraries such as scipy.integrate.odeint package. To reduce the number of free variables, we assume the qubit is in thermal equilibrium with an effective qubit temperature $T_{\mathrm{qb}}$. This allows us to express the upward excitation rates in terms of the downward relaxation rates: $\Gamma_{1}^{i \rightarrow j} = \exp(-\hbar \omega_{ij}/k_{B}T_{\mathrm{qb}}) \Gamma_{1}^{j \rightarrow i}$ for $E_{j} > E_{i}$. By fitting the experimental result to the numerical solution of \eref{eqn:three_state_decay_ode}, we can extract the individual relaxation rates and the effective qubit temperature.

\subsection{Numerical simulation of erasure check}

While the ODE approach is efficient for fitting ensemble averages, it is less suited for simulating mid-circuit erasure checks. The ODEs track continuous probabilities, so they cannot easily capture the discrete digital nature of erasure detection events and the subsequent post-selection process.

Thus, we adopted a Monte Carlo (MC) simulation to model the erasure check. In this framework, the system is modeled as occupying exactly one of the three states at any given time step. Transitions between states are determined stochastically based on probabilities derived from the relaxation rates $\Gamma_{1}^{i \rightarrow j}$. This "shot-by-shot" approach allows us to explicitly simulate the mid-circuit detection of $|\mathrm{e} \rangle$ state and more faithfully model the effect of discarding specific shots.

In the Monte Carlo framework, a successful erasure detection is defined by the system dwelling in the $|\mathrm{e}\rangle$ state for a cumulative duration exceeding a specified threshold fraction of the erasure-check duration. This is to approximately models the experimental condition, where the erasure check is imperfect and the duration is finite. In addition, we incorporated the possible change of decay rates ($|\mathrm{e}\rangle \rightarrow |g\rangle$ and $\mathrm{|f\rangle \leftrightarrow |e\rangle}$) observed in our experimental $T_{1}$ measurements when the erasure-check signal is active. The specific implementation steps are detailed in Algorithm \ref{alg:mc_algorithm}. Notably, the Monte Carlo simulation successfully reproduced the non-exponential decay observed in our experimental $T_{1}$ measurements following the postselection against the erasure error events. In particular, we found this method useful to estimate $T_{1}^{\mathrm{f} \rightarrow \mathrm{g}}$, as the simulation result after discarding erasure events becomes more sensitive to this parameter compared to the three-state decay model. This agreement indicates the model captures the dominant population dynamics observed in the g-f encoded integer fluxonium.

\begin{algorithm}
    \caption{Monte-Carlo simulation with erasure-check during evolution of three states}
    \label{alg:mc_algorithm}
    \begin{algorithmic}
        \State $\mathit{prob\_ij\_base}$: Transition probability from state $i$ to $j$ when erasure check is not present. 
        \State $\mathit{prob\_ij\_ec}$: Transition probability from state $i$ to $j$ when erasure check is turned on.
        \State $\mathit{threshold}$: Minimum erasure detection count during erasure check duration to flag as an erasure event.
        \For{$\mathit{shots}$ in $1,2,\ldots, N$}
            \State $\mathit{flag} \leftarrow \text{False}$, $\mathit{ec\_count} \leftarrow 0$, $\mathit{state} \leftarrow \mathit{init\_state}$ 
            \For{$\mathit{t\_idx}$ in $0,1,\ldots, M$}
                \State $\mathit{echeck} \leftarrow 0$
                \State $r \leftarrow \text{random value between 0 and 1}$
                \If{$\mathit{t\_idx} \pmod{\mathit{ec\_cadence}} < \mathit{ec\_duration}$}
                    \State $\mathit{echeck} \leftarrow 1$ 
                \EndIf
                \If{$\mathit{echeck} = 1$}
                    \State $\mathit{prob\_ij} \leftarrow \mathit{prob\_ij\_ec}$
                \EndIf
                \If{$\mathit{echeck} = 0$}
                    \State $\mathit{prob\_ij} \leftarrow \mathit{prob\_ij\_base}$
                \EndIf
                \If{$\mathit{state}$ is $f$}
                    \If{$r < \mathit{prob\_fe}$}
                        \State $\mathit{state} \leftarrow e$
                    \ElsIf{$\mathit{prob\_fe} < r < \mathit{prob\_fe} + \mathit{prob\_fg}$}
                        \State $\mathit{state} \leftarrow g$
                    \EndIf
                \ElsIf{$\mathit{state}$ is $e$}
                    \If{$\mathit{echeck} = 1$}
                        \State $\mathit{ec\_count} \leftarrow \mathit{ec\_count} + 1$
                    \EndIf
                    \If{$r < \mathit{prob\_ef}$}
                        \State $\mathit{state} \leftarrow f$
                    \ElsIf{$\mathit{prob\_ef} < r < \mathit{prob\_ef} + \mathit{prob\_eg}$}
                        \State $\mathit{state} \leftarrow g$
                    \EndIf
                \ElsIf{$\mathit{state}$ is $g$}
                    \If{$r < \mathit{prob\_ge}$}
                        \State $\mathit{state} \leftarrow e$
                    \ElsIf{$\mathit{prob\_ge} < r < \mathit{prob\_ge} + \mathit{prob\_gf}$}
                        \State $\mathit{state} \leftarrow f$ 
                    \EndIf
                \EndIf 
                \If{$\mathit{t\_idx} \pmod{\mathit{ec\_cadence}} = 0$ \textbf{and} $\mathit{t\_idx} > 0$}
                    \If{$\mathit{ec\_count} > \mathit{threshold}$}
                        \State $\mathit{flag} \leftarrow \text{True}$
                    \EndIf 
                    \State $\mathit{ec\_count} \leftarrow 0$
                \EndIf
            \EndFor
            \If{$\mathit{flag}$ is False} keep the shot
            \EndIf
            \If{$\mathit{flag}$ is True} discard the shot
            \EndIf
        \EndFor
    \end{algorithmic}
\end{algorithm}

\begin{figure}[H]
    \centering
    \includegraphics[width=0.99\linewidth]{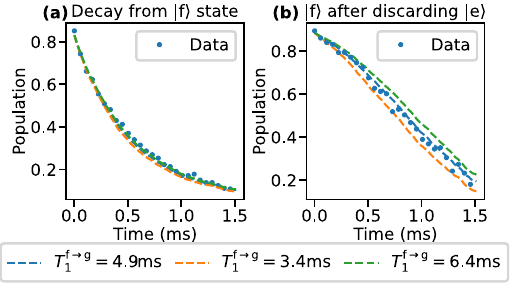}
    \caption{Monte Carlo simulations of the $|f\rangle$ state population as a function of the direct relaxation rate $T_{1}^{\mathrm{f \to g}}$. (a) Simulated population decay from the $\mathrm{|f\rangle} $ state. (b) $\mathrm{|f\rangle}$ state population after discarding the $|\mathrm{e\rangle}$ state population. The simulated curves demonstrate the sensitivity of the decay profile to $T_1^{\mathrm{f \to g}}$ values of 3.4 ms, 4.9 ms, and 6.4 ms. By filtering out erasure-state events, the resulting decay curves exhibit higher sensitivity to $T_1^{\mathrm{f \to g}}$ variations than (a), allowing for a more rigorous characterization of the direct relaxation within the computational subspace.}
    \label{fig:monte_carlo_example}
\end{figure}

\section{\label{app:parameter_regime} PARAMETER SPACE OF AN INTEGER FLUXONIUM AS A g-f ERASURE QUBIT}
\begin{figure} [H]
    \centering
    \includegraphics[width=0.99 \linewidth]{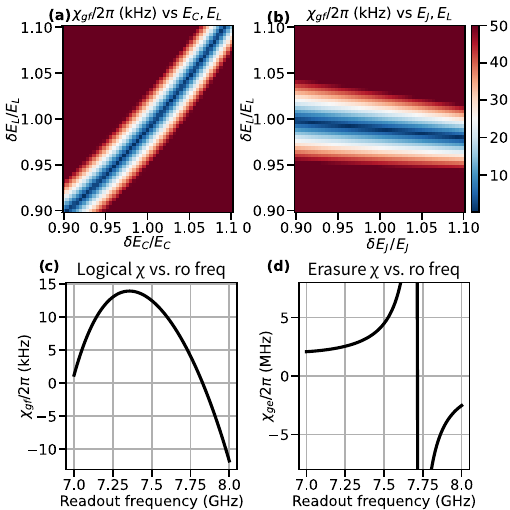}
    \caption{(a, b) Simulated dispersive shift magnitude of the computational subspace ($|\chi_{\mathrm{gf}}|/2\pi$) as a function of fluxonium parameters. Variations are shown within a $\pm$10\% range relative to the base values of $E_{C}, E_{J},E_{L}$. In (a), $E_{J}$ is fixed while in (b), $E_{C}$ is fixed. The color scale is saturated at \SI{50}{kHz} to highlight the target regime for high-fidelity erasure checks. (b) demonstrates that $\chi_{\mathrm{gf}}$ is significantly less sensitive to variations in $E_{J}$ compared to $E_{C},E_{L}$. (c, d) $\chi_{\mathrm{gf}}/2\pi$, $\chi_{\mathrm{ge}}/2\pi$ vs readout frequency. $|\chi_{\mathrm{ge}}|/2\pi$ is over \SI{2}{MHz} while $\chi_{\mathrm{gf}}$ is limited to $\pm$\SI{15}{kHz} within the readout frequency of 7.0--\SI{8.0}{GHz}. }
    \label{fig:parameter_space}
\end{figure}

Integer fluxonium regime is defined by the condition $2 \pi^{2} E_{L} \ll \sqrt{8 E_{J} E_{C}}$. To eliminate the need for an ancilla qubit, an additional condition must be satisfied: $\chi_{\mathrm{gf}} \ll \chi_{\mathrm{ge}}$. Based on numerical simulations, we observed that $\chi_{\mathrm{gf}} $ can be suppressed neaer zero while $\chi_{\mathrm{ge}}$ remains on the order of several megahertz, thereby providing a favorable condition for fast, high-fidelity erasure checks. We further assessed the robustness of $\chi_{\mathrm{gf}} $ by sweeping fluxonium qubit parameters $E_{C}, E_{J}, E_{L}$ and readout resonator frequency $f_{\mathrm{res}}$. 

For these simulations, we established base qubit parameters of $E_{C}/h = 1.35$ GHz, $E_{J}/h = 4.5$ GHz, and $E_{L}/h = 0.19$ GHz, with a readout frequency of $7.4$ GHz and a qubit-resonator coupling strength of $g/2\pi = 0.1$ GHz, where $g$ is modeled by the Hamiltonian $\hat{H}_{\mathrm{qb-res}} = g \hat{n}_{\mathrm{qb}} (\hat{a}^\dagger + \hat{a})$. Under these conditions, the dispersive shifts are $\chi_{\mathrm{ge}}/2 \pi = 3.324$ MHz and $\chi_{\mathrm{gf}}/2\pi = 13.77$ kHz.

Next, we swept each parameter by $\pm 10\%$ [\fref{fig:parameter_space}(a, b)] and calculated $\chi_{\mathrm{gf}}$. We observed that $\chi_{\mathrm{gf}}$ is roughly equally sensitive to variations in $E_{C}$ and $E_{L}$. To maintain $|\chi_{\mathrm{gf}}|/2\pi < 50$ kHz, the feasible region constitutes approximately 20\% of the explored parameter space. In contrast, $\chi_{\mathrm{gf}}$ is considerably less sensitive to changes in $E_{J}$ than to $E_{C}$ or $E_{L}$. While the tolerances for $E_{C}$ and $E_{L}$ are relatively stringent ($\pm 4\%$), they remain within the reach of modern fabrication precision.

Additionally, we calculated $\chi_{gf}$ and $\chi_{ge}$ while sweeping the readout resonator frequency. Across a range of $7$--$8$ GHz, $\chi_{gf}/2\pi$ remained within $\pm 15$ kHz, while $\chi_{ge}/2\pi$ was maintained above $2$ MHz. This stability held across the entire range, except near the pole associated with the avoided crossing between the fluxonium transition and the resonator mode.

\section{\label{app:effect_of_leakage_on_T2} INCOHERENT ERROR CONTRIBUTION FROM LEAKAGE IN A g-f ERASURE QUBIT}

In this section, we calculate the effect of leakage error $T_{1}^{\mathrm{f \rightarrow e}}$ to the $T_{2}$ and gate fidelity that we measure. The Kraus operators for this quantum channel can be expressed as follows.
\begin{align}
    E_{0} = \begin{bmatrix}
        1 & 0  \\
        0 & \sqrt{1 - p} \\ 
        0 & 0 
    \end{bmatrix},~E_{1} = \begin{bmatrix}
        0 & 0 \\ 0 & 0 \\ 0 & \sqrt{p} 
    \end{bmatrix}
\end{align}
where the first two rows represent the two computational states and the third row corresponds to the erasure state $|\mathrm{e} \rangle$. $p$ is the probability when the initial state was in $|\mathrm{f}\rangle$, the state will be in $|\mathrm{e}\rangle$, thus, $p = 1 - \exp(-t/T_{1}^{\mathrm{fe}})$. We can confirm that $E_{0}^{\dagger} E_{0} + E_{1}^{\dagger} E_{1}$ is an identity matrix as a sanity check. In the ordered output basis $\mathrm{|g\rangle, |f\rangle, |e\rangle}$, the channel maps a $2 \times 2$ computational-subspace density matrix to the following $3 \times 3$ density matrix.
\begin{align}
    \sum_{k=0}^{1} E_{k} \rho E_{k}^{\dagger} = \begin{bmatrix}
        \rho_{00} & \sqrt{1-p} \rho_{01} & 0 \\ 
        \sqrt{1-p} \rho_{10} & (1-p) \rho_{11} & 0 \\
        0 & 0 & p \rho_{11}
    \end{bmatrix}
\end{align}
where $\rho_{ij}$ is the matrix element of $i$th row, $j$th column of $\rho$. In a $T_{2}$ sequence, we measure the off-diagonal element $\rho_{01}$, which satisfies $\rho_{01}(t) = \sqrt{1-p } \rho_{01} = \exp(-t/2T_{1}^{\mathrm{fe}}) \rho_{01}$. Thus, the effect of leakage error contributes half of its rate to the $1/T_{2}$. 

Next, we calculate the effect of $\mathrm{f \rightarrow e}$ leakage error on randomized benchmarking. Given the Kraus operators $E_{k}$, the gate fidelity satisfies the following equation. 
\begin{align}
    F = \frac{1}{d(d+1)} \left[\sum_{k}\mathrm{Tr}[M_{k}M_{k}^{\dagger}] + |\mathrm{Tr}[M_{k}]|^{2} \right]
\end{align}
where $M_{k} = PE_{k}$. $P=\mathrm{|g\rangle \langle g|+ |f\rangle \langle f|}$ is the projection operator to the computational subspace. This makes $M_{0}$ the only non-zero matrix. Thus, we have the following equation.
\begin{align}
\begin{split}
    &\mathrm{Tr}[M_{0}M_{0}^{\dagger}] = 2 - p,~|\mathrm{Tr}[M_{0}]|^{2} = 2-p + 2\sqrt{1-p} \\
    &\Rightarrow~ F = \frac{1}{6} (4-2p + 2 \sqrt{1-p})
\end{split}
\end{align}
Substituting $p = 1-\exp(-t/T_{1}^{\mathrm{fe}})\approx t/T_{1}^{\mathrm{fe}}$ and $\sqrt{1-p} \approx 1-p/2$, we obtain:
\begin{align}
F \approx \frac{1}{6} (6 - 3p)= 1-\frac{t}{2 T_{1}^{\mathrm{fe}}}   
\end{align}
Compared to usual error contribution of $t/3T_{1}$ within the computational subspace, in single-qubit gate, leakage error contributes larger amount.

\section{\label{app:lrb_protocol} LEAKAGE RANDOMIZED BENCHMARKING PROTOCOL}

In this section, we provide the detailed procedure of extracting the leakage error from the randomized benchmarking result, following~\cite{Wood2018_LRB}. Define $i_{m}$ as the Clifford sequence of length $m$ and $\mathcal{R}_{m+1}$ to be the recovery Clifford for sequence $i_{m}$, which recovers the result to the ground state $|\mathrm{g}\rangle $. That is, we define as the following.
\begin{align}
    i_{m} = \mathcal{C}_{m} \circ \mathcal{C} _{m-1} \circ \cdots \circ \mathcal{C}_{1},~ \mathcal{R}_{m+1} \circ i_{m} = \mathcal{I} 
\end{align}

Also, let $p_{\mathrm{comp}}(m; i_{m})$ to be the probability that the final state would stay in the computational state after applying Clifford sequence $i_{m}$ followed by $\mathcal{R}_{m+1}$. Similarly $p_{\mathrm{0}}(m;i_{m})$ can be defined as the probability that the final state would be measured at the ground state. In the experiment, since we take the ensemble average over multiple random Clifford sequences, we define $p_{\mathrm{comp}}(m) \equiv \mathrm{E}_{i_{m}}[p_{\mathrm{comp}}(m;i_{m})]$ and $p_{\mathrm{0}}(m) \equiv \mathrm{E}_{i_{m}}[p_{\mathrm{0}}(m;i_{m})]$. Then, the leakage RB protocol proceeds as the following.
\begin{itemize}
    \item The computational space population, $p_{\mathrm{comp}}(m)$, is fitted to a single-exponential decay $A + B \lambda_{1}^{m}$ to extract the leakage error rate per Clifford gate $L_{1} = (1 - A)(1 - \lambda_{1})$.
    \item We fit the ground state population $p_{\mathrm{0}}(m)$, to a double-exponential decay, $A_{0} + B_{0} \lambda_{1}^{m} + C_{0} \lambda_{2}^{m}$. To mitigate the inherent instability of multi-exponential fitting, we constrain $A_{0}$ by first fitting the $p_{0}(m)$ to a single-exponential decay $p_{\mathrm{0}}(m) = A_{0} +C_{0}'\lambda_{L}^{m}$, under the physically motivated assumption that both models should converge to the same steady-state population as $m\rightarrow \infty$. 
    \item From the double-exponential fit, the average fidelity per Clifford can be estimated by 
    \begin{align}
        F = \frac{1}{d} [(d - 1)\lambda_{2} + 1 - L_{1}]
    \end{align} where $d=2$ for the single qubit.
    \item To calculate the average fidelity per $\sqrt{X}$, we divide the total error rate per Clifford by the average number of $\sqrt{X}$ gates per Clifford, which is 2.208. This value reflects our specific implementation, where $\pi$-pulses are constructed from two $\pi/2$ pulses, as detailed in \tref{tab:clifford_1qb_implementation}.
\end{itemize}

\begin{figure}[H]
    \centering
    \includegraphics[width=0.99\linewidth]{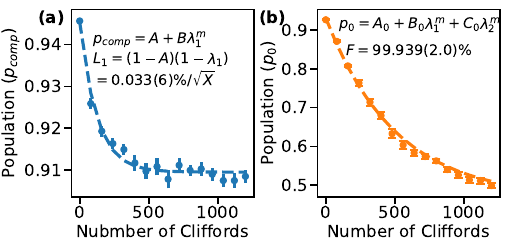}
    \caption{(a) Population fraction remaining the computational space ($p_{\mathrm{comp}}$) by number of Clifford gates in RB sequence. By fitting the data to a single-exponential decay, we obtained leakage error per $\sqrt{X}$ ($L_{1}$) of 0.033(6)\%. (b) Population measured in the ground state $|\mathrm{g}\rangle$ after $m $ Clifford gates in RB sequence ($p_{0}(m)$). We estimate gate fidelity of 99.939(2)\% per $\sqrt{X} $ from the double-exponential fit.}
    \label{fig:lrb_result}
\end{figure}

The measured data to obtain leakage error $L_{1}$ and fidelity $F$ is shown in \fref{fig:lrb_result}.

\begin{table}[H]
    \centering
    \begin{tabular}{c|c|c}
        \hline
        Clifford & Implementation & Gate count \\
        \hline
        I &  & 1 \\
        ${X_{2p},X_{2m},Y_{2p},Y_{2m}}$  &  & 4 \\
        $X_{p}$& $X_{2p}X_{2p}$ & 2\\
        $Y_{p}$ & $Y_{2p}Y_{2p}$ & 2\\
        $Z_{p}$ & $X_{2p}X_{2p}Y_{2p}Y_{2p}$ & 4 \\
        $Y_{2p}X_{2p}$ & & 2 \\
        $Y_{2m}X_{2p}  $ & & 2 \\
        $ Y_{2p}X_{2m} $ & & 2 \\
        $ Y_{2m}X_{2m} $ & & 2 \\
        $ X_{2p}Y_{2p} $ & & 2 \\
        $ X_{2m}Y_{2p} $ & & 2 \\
        $ X_{2p}Y_{2m} $ & & 2 \\ 
        $  X_{2m}Y_{2m}$ & & 2 \\ 
        $Y_{2p}X_{p}$ & $X_{2p}X_{2p}Y_{2p}$ & 3 \\
        $Y_{2m}X_{p}$ & $Y_{2m}X_{2p}X_{2p}$ & 3 \\
        $X_{2p}Y_{p}$ & $X_{2p}Y_{2p}Y_{2p}$ & 3 \\
        $X_{2m}Y_{p}$ & $X_{2m}Y_{2p}Y_{2p}$ & 3 \\
        $X_{2p} Y_{2p} X_{2m}$ &&3 \\
        $X_{2p}Y_{2m}X_{2m}$ && 3\\
        $X_{2p}Y_{2p} X_{2p}$ && 3\\
        $X_{2m}Y_{2p}X_{2m}$ && 3 \\
        \hline
        & Total & 53 \\
        \hline
    \end{tabular}
    \caption{Implementation of single-qubit Cliffords in the paper. The total number of gates is 53 for 24 Cliffords, thereby resulting in 53/24=2.208 gates per Clifford. The implementation is left blank if the type of Clifford and the implementation is the same form.}
    \label{tab:clifford_1qb_implementation}
\end{table}

\begin{widetext}
    
\section{\label{app:twophoton_drive} THEORY OF TWO-PHOTON DRIVE}
Since the two logical states $\mathrm{|g\rangle,|f\rangle}$ have the same parity in an integer fluxonium, single-photon matrix element between them is suppressed. To perform single-qubit gates within this computational subspace, we employed a two-photon drive mediated by the intermediate $|\mathrm{e} \rangle$ state. In this section, we derive the effective Rabi amplitude of the two-photon drive and its implication for gate calibration. Consider a generic three-level system described by the following Hamiltonian.
\begin{align}
    \hat{H}/\hbar  = \omega_{\mathrm{e}}\mathrm{|e\rangle \langle e|} + \omega_{\mathrm{f}} \mathrm{|f\rangle \langle f|} + \Omega_{d} \cos (\omega_{d}t - \phi) ({O}_{\mathrm{ge}}\mathrm{|g\rangle \langle e| } + O_{\mathrm{ef}} \mathrm{|e\rangle \langle f|} + \mathrm{H.c.})
\end{align}
where the ground state energy is defined as zero and $\Omega_{d}$ represents the drive amplitude. For two-photon drive implementation, usually $2\omega_{d} \approx \omega_{\mathrm{f}}$, but we do not need this condition until later. We move the Hamiltonian in the rotating frame of reference of $\omega_{d}$ by applying $\hat{U} = \exp(i\omega_{d}t\mathrm{|e\rangle \langle e|} + i 2\omega_{d}t \mathrm{|f \rangle \langle f|} )$ and calculate Hamiltonian in the rotating frame $\hat{H}_{R} = \hat{U} \hat{H} \hat{U}^{\dagger} - i \hbar \hat{U} (d\hat{U}^{\dagger}/dt)$. After the calculation, if we remove all the fast-oscillating terms $\exp(\pm i 2 \omega_{d}t)$ (Rotating-Wave Approximation), we get the following time-independent form.
\begin{align}
    \hat{H}_{R}/\hbar \approx  \Delta_{e} \mathrm{|e\rangle \langle e|} + \Delta_{\mathrm{f}} \mathrm{|f\rangle \langle f|} + \frac{\Omega_{d}}{2} (O_{\mathrm{ge}} e^{-i \phi } \mathrm{|g \rangle \langle e|} + O_{\mathrm{ef}} e^{-i \phi }\mathrm{|e\rangle  \langle f|} + \mathrm{H.c.})
\end{align}
where $\Delta_{\mathrm{e}} = \omega_{\mathrm{e}} - \omega_{d}$ and $\Delta_{\mathrm{f}}=\omega_{\mathrm{f}}- 2\omega_{d} $. Then, we solve the Schr\"odinger equation in the rotating frame, first by writing $|\psi(t)\rangle = c_{\mathrm{g}} \mathrm{|g\rangle } +c_{\mathrm{e}} \mathrm{|e\rangle} + c_{\mathrm{f}} \mathrm{|f\rangle}$.
\begin{align}
\begin{split}
    i \dot{c}_{\mathrm{f}} &= \Delta_{\mathrm{f}} c_{\mathrm{f}} + \frac{\Omega_{d}}{2} O_{\mathrm{fe}} e^{i \phi } c_{\mathrm{e}} \\ 
    i \dot{c}_{\mathrm{e}} & = \Delta_{\mathrm{e}}c_{\mathrm{e}} + \frac{\Omega_{d}}{2} \left( O_{\mathrm{ef}} e^{-i \phi } c_{\mathrm{f}} + O_{\mathrm{eg}} e^{i \phi }  c_{\mathrm{g}}\right) \\ 
    i \dot{c}_{\mathrm{g}} & = \frac{\Omega_{d}}{2} O_{\mathrm{ge}} e^{-i \phi } c_{e}
\end{split}
\end{align}

Now we assume $|\Delta_{e}| \gg |\Omega_{d}O_{\mathrm{ge}}|,|\Omega_{d} O_{\mathrm{ef}}|$, then $c_{e}$ remains small and therefore we can apply adiabatic elimination which we set $\dot{c}_{e} = 0$. Then, we can substitute $c_{e} = -\Omega_{d} \left( O_{\mathrm{ef}} e^{-i \phi } c_{\mathrm{f}} + O_{\mathrm{eg}} e^{i \phi }  c_{\mathrm{g}}\right)/2 \Delta_{\mathrm{e}}$, and reduce the three-level Hamiltonian into an effective two-level system Hamiltonian consisting of $\mathrm{|g\rangle, |f\rangle}$.
\begin{align}
    \begin{bmatrix}
        i\dot{c}_{\mathrm{f}} \\ i\dot{c}_{\mathrm{g}}
    \end{bmatrix} = \begin{bmatrix}
        \Delta_{\mathrm{f}} - \frac{\Omega_{d}^{2}}{4 \Delta_{\mathrm{e}}} |O_{\mathrm{ef}}|^{2} & -\frac{\Omega_{d}^{2}}{4 \Delta_{e}} O_{\mathrm{eg}} O_{\mathrm{fe}} e^{i 2\phi }\\ -\frac{\Omega_{d}^{2} }{4 \Delta_{e}}O_{\mathrm{ge}} O_{\mathrm{ef}} e^{-i 2 \phi }  & - \frac{\Omega_{d}^{2}}{4 \Delta_{e}} |O_{\mathrm{ge}}|^{2}  
    \end{bmatrix} \begin{bmatrix}
        c_{\mathrm{f}} \\ c_{\mathrm{g}}
    \end{bmatrix}. 
\end{align}
The diagonal terms represent the drive-induced stark shift of the two levels. The off-diagonal terms indicate the effective Rabi amplitude. There are two facts worth noticing. One, Rabi amplitude in two-photon drive is proportional to $\Omega_{d}^{2} $ instead of $\Omega_{d}$. This actually has an important implication such that the effect of envelope of the drive pulse is effectively squared to the Rabi oscillation, meaning that if we want the cosine-shaped ramp up and down effectively, we need square-root-cosine shaped envelope for the input. Next, note that there is a phase term $e^{i 2\phi}$. This indicates the $\phi$-phase shift in microwave drive results in $2\phi$ phase shift in the effective result. Therefore, if we define $X$ gate to be prepared at $\phi=0$, to implement $Y$ gate one has to shift the phase by $\pi/4$ instead of $\pi/2$ because of the $2\phi$ term.

\end{widetext}

\section{\label{app:nbar_calibration} AVERAGE READOUT PHOTON NUMBER CALIBRATION}

\begin{figure}[H]
    \centering
    \includegraphics[width=0.99\linewidth]{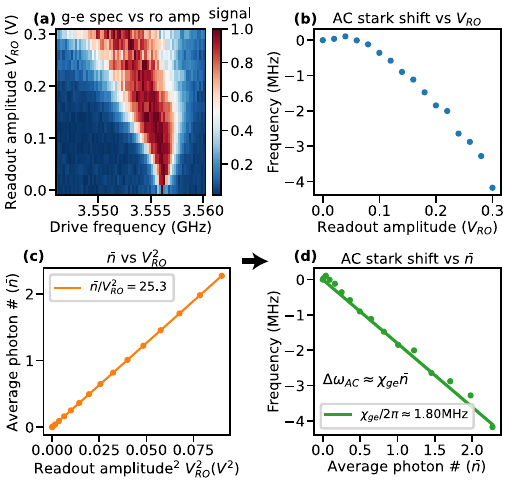}
    \caption{(a) Two-tone spectroscopy measurement of g-e transition while the readout signal was turned on. We swept the readout signal amplitude and measured the transition frequency shift (AC Stark shift). (b) AC Stark shift as a function of readout amplitude. (c) We estimated the relation between the readout amplitude squared ($V_{RO}^{2}$) and average photon number in the readout resonator $\bar{n}$ by comparing the theoretically calculated AC Stark shift and the measured data. We obtained the best fit when $\bar{n}/V_{RO}^{2}=25.3~\mathrm{V}^{-2}$ . (d) We replotted the AC Stark shift as a function of average photon number $\bar{n}$. We could estimate $\chi_{\mathrm{ge}}/2\pi=$\SI{1.80}{MHz} from the relation $\Delta \omega_{\mathrm{AC}} =\chi_{\mathrm{ge}} \bar{n}$ where $\omega_{\mathrm{AC}}/2\pi$ is the AC Stark shifted frequency.}
    \label{fig:nbar_calibration}
\end{figure}

We calibrated the average photon number ($\bar{n}$) in the readout resonator by measuring the AC Stark shift~\cite{Schuster2005_acstarkshift} of the g--e transition. We performed two-tone spectroscopy of g--e transition while the readout signal frequency of \SI{7.3894}{GHz} was turned on and its amplitude was swept [\fref{fig:nbar_calibration}(a)]. We plotted the qubit frequency as a function of the square of the readout amplitude [\fref{fig:nbar_calibration}(b)]. Next, we fit the data to find the correct ratio between the average photon number and the readout amplitude squared ($V_{RO}^{2}$). Specifically, we calculated the qubit-resonator system with the resonator at the coherent state of amplitude $\sqrt{\bar{n}}$, and calculated qubit frequency shift compared to the measured value. From the theoretical fit, we found $\bar{n}/V_{\mathrm{RO}}^{2} = 25.3$ provided the best agreement with the measured data. Based on the fit result, we replotted the AC Stark shift as a function of $\bar{n}$ [\fref{fig:nbar_calibration}(d)]. 

From the calibration result, we estimate the photon number during the erasure check was around $\bar{n}\approx 1.58$ ($V_{\mathrm{RO}}=$\SI{0.25}{V}) for cycle 2. For cycle 1, the erasure-check amplitude was \SI{0.28}{V}, however, due to the different resonant frequency and the erasure-check frequency, the actual corresponding average photon number was calculated to be $\bar{n} \approx 1.7$.

\section{\label{app:gate_calibration_procedure} SINGLE-QUBIT CALIBRATION PROCEDURE}

\begin{widetext}
\begin{figure}[b]
    \centering
    \includegraphics[width=0.8\linewidth]{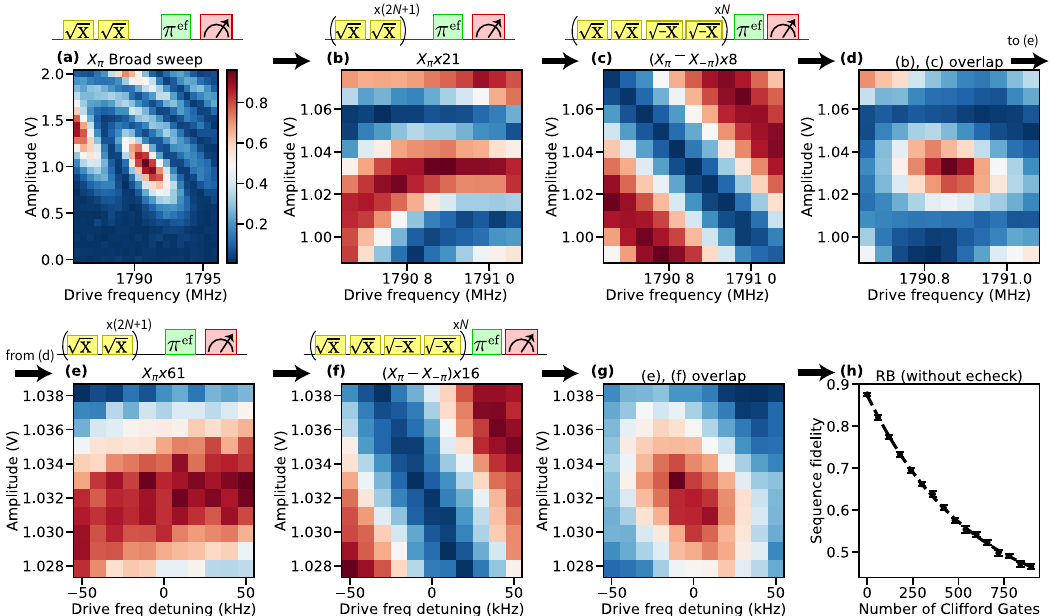}
    \caption{Single-qubit gate calibration procedure in the integer fluxonium in this research. (a) Broad sweep of drive amplitude and drive frequency using a single $\pi$ pulse to find an initial calibration point. (b) Drive amplitude, frequency calibration using odd numbers of $\pi$-pulses. (c) Drive amplitude, frequency calibration using pulse train of $X_{\pi}-X_{-\pi}$. (d) Overlap of result (b) and (c). We calculate (1 - result from (c)) and multiply it the result from (b). (e) Finer calibration of drive amplitude and frequency using 61 $\pi$-pulses. (f) Finer calibration of drive amplitude and frequency using 16 $X_{\pi} - X_{-\pi}$ pairs. (g) Overlap of result (e) and (f), the overlap was calculated similar to panel (d). (h) RB trace after single-qubit gate calibration.}
    \label{fig:1qb_gate_calibration}
\end{figure}
\end{widetext}

We calibrated the single-qubit gates of the integer fluxonium implemented by a two-photon microwave drive, similar to the procedure introduced in~\cite{Schirk2025_flxnm_multiphoton}. The optimal drive frequency and amplitude were determined using error-amplifying pulse sequences designed to isolate over-rotation and frequency detuning errors. Due to a substantial AC Stark shift, minor miscalibrations in drive amplitude induced significant frequency shifts; consequently, both parameters were calibrated simultaneously through an iterative optimization process to ensure high-fidelity operations. The entire calibration procedure is summarized in \fref{fig:1qb_gate_calibration}.

\section{\label{app:echeck_vz_calibration} ERASURE-CHECK INDUCED SINGLE-QUBIT PHASE CALIBRATION}

\begin{figure}[t]
    \centering
    \includegraphics[width=0.95\linewidth]{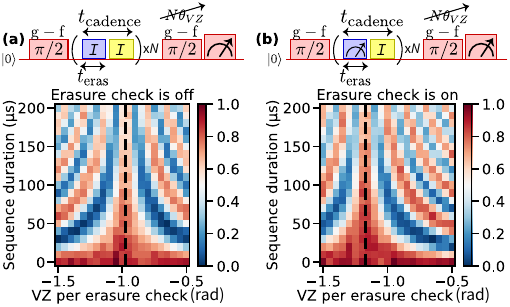}
    \caption{(a) Pulse sequence for the virtual-Z (VZ) angle calibration and its result when erasure-check is off. Erasure-check cadence $t_{\mathrm{cadence}}$ was \SI{10}{\micro s} and the erasure-check duration $t_{\mathrm{eras}}$ was \SI{1.1}{\micro s}. We obtained an additional single-qubit phase (VZ angle) accumulation of \SI{-0.975}{rad} per erasure check cadence. (b) Pulse sequence for the VZ angle calibration and its result when erasure check was applied. We obtained VZ angle of \SI{-1.170}{rad} per erasure check cadence.}
    \label{fig:ec_induced_vz}
\end{figure}

In the integer fluxonium we measured, each erasure check induces qubit frequency shift proportional to $\chi_{\mathrm{gf}}$. Since $\chi_{\mathrm{gf}}$ is small but nonzero, we have to apply a virtual-$Z$ frame update to all the single-qubit gates following each erasure check to compensate for this shift. To calibrate the amount of phase induced by each erasure check, we measured a Ramsey-like sequence while applying virtual-Z gate to the second $\pi/2$ pulse proportional to the evolution time [\fref{fig:ec_induced_vz}].

Without erasure check (by setting erasure-check amplitude to zero), we calibrated compensation angle of \SI{-0.975}{rad} per \SI{10}{\micro s} erasure check cadence to cancel the single-qubit phase accumulation during the free evolution. This is due to the drive-induced frequency detuning from the bare g-f transition frequency during the idle periods. Once the erasure-check was turned on, the calibrated compensation angle changed from \SI{-1.170}{rad} per erasure check cadence. Thus, we can estimate the additional single-qubit phase accumulation from the erasure check was \SI{-0.195}{rad}. 

In the randomized benchmarking experiment, we applied erasure check of \SI{1.1}{\micro s} duration followed by \SI{1}{\micro s} free evolution time to allow resonator-photon ringdown. Thus, the amount of single-qubit phase accumulated during this timeline is \SI{-0.975}{rad}$\times $\SI{2.1}{\micro s}/\SI{10}{\micro s} + \SI{-0.195}{rad} = \SI{-0.3998}{rad}. Therefore, for each erasure check, we add virtual-Z gate with angle of \SI{-0.3998}{rad} for all Clifford gates following an erasure check.

\section{\label{app:T1_reduction_during_echeck} $T_{1}$ DEGRADATION DURING AN ERASURE CHECK}

We measured the $T_{1}$ between $|\mathrm{e} \rangle$ and $\mathrm{|g \rangle}$ during the erasure check as we swept its amplitude and frequency. We initialized the system in the erasure state $\mathrm{|e \rangle}$ and subsequently measured the ground state population as a function of time. Next, we replotted the result to analyze the correlation with the average photon number in the readout resonator, $\bar{n}$ using the photon number calibration result. We measured a significant drop in the $T_{1}^{\mathrm{e \rightarrow g}}$ from maximum of \SI{300}{\micro s} without erasure check to below \SI{20}{\micro s} when $\bar{n} \geq 1.0$. The underlying mechanisms for this coherence drop are not identified by this study, but one possible mechanism is increased coupling to two-level systems resulting from readout-induced spectral broadening, described in~\cite{Huang2026ro_induced_t1, Thorbeck2024_ro_induced_t1}. However, the present data do not identify the microscopic origin. In contrast, we observed no statistically significant reduction in $T_{1}^\mathrm{f \rightarrow e}$ during the erasure check.
\begin{figure}[H]
    \centering
    \includegraphics[width=0.99\linewidth]{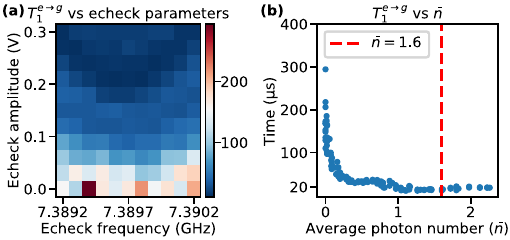}
    \caption{(a) $T_{1}^{\mathrm{e \rightarrow g}}$ as a function of erasure-check frequencies and amplitudes. (b) Replotting of (a) as a function of average photon number $\bar{n}$ in the readout resonator. We observed a sharp drop in $T_{1}$ from \SI{300}{\micro s} to below \SI{20}{\micro s} when the erasure check was turned on.}
    \label{fig:echeck_mist}
\end{figure}

\onecolumngrid
\end{document}